\documentclass[11pt,aps,prd,preprint,superscriptaddress]{revtex4-1}
\usepackage{amsmath}
\usepackage[dvips]{graphicx}
\usepackage[english]{babel}
\newcommand{\be}{\begin{equation}}
\newcommand{\ee}{\end{equation}}
\newcommand{\ben}{\begin{eqnarray}}
\newcommand{\een}{\end{eqnarray}}

\begin{document}
\title{A thermodynamic characterization of future singularities?}
\author{Diego Pav\'{o}n\footnote{E-mail: diego.pavon@uab.es}}
\affiliation{Department de F\'{\i}sica, Universitat Aut\`{o}noma
de Barcelona, 08193 Bellaterra (Barcelona), Spain}
\author{Winfried Zimdahl\footnote{E-mail: winfried.zimdahl@pq.cnpq.br}}
\affiliation{Universidade Federal do Esp\'{\i}rito Santo,
Departamento
de F\'{\i}sica, Av. Fernando Ferrari, 514,\\
Campus de Goiabeiras, CEP 2905-910, Vit\'{o}ria, Esp\'{\i}rito
Santo, Brasil}
\begin{abstract}
In this Letter we consider three future singularities in different
Friedmann-Lema\^{\i}tre-Robertson-Walker scenarios and show that
the universe departs more and more from thermodynamic equilibrium
as the corresponding singularity is approached. Though not proven
in general, this feature may characterize future singularities of
homogeneous and isotropic cosmologies.\\
\[\]
\underline{Key words}: {\em mathematical cosmology, singularities,
thermodynamics}
\end{abstract}

 \maketitle

\section{Introduction}\label{introduction}
\noindent Cosmological singularities arise as displeasing features
in mathematical models of the universe \cite{stephen-george,Bob};
world lines terminate and/or physical quantities diverge. It is
usually argued that this results because of the exceedingly
simplicity of the models in question, implying that when we reach
a much better understanding of the physical processes that take
place under the most extreme conditions we will be able to design
realistic models free of singularities.
\  \\

\noindent Perhaps the most worrying of all is the Big Bang
singularity which persists even if the standard cosmological model
is corrected with the addition of an era of inflationary expansion
just before the radiation dominated epoch. This explains the
interest raised by proposals of universes, such as  bouncing
\cite{bouncing} and emergent \cite{emergent}, with no beginning at
all.
\  \\

\noindent Singularities are heralded, among other things, by the
growing without bound of key physical quantities, such as energy
densities and pressures. Here we focus on future cosmic
singularities in Friedmann-Lema\^{\i}tre-Robertson-Walker (FLRW) universes and
ask ourselves if they may be characterized by some thermodynamic
distinctive feature. Our provisory answer is in the affirmative.
We reach this tentative conclusion after considering three
singularities, linked to different cosmological models. As it
turns out in the three cases the total entropy of the systems does
not tend to a maximum as the singularity is approached. That is to
say, the systems do not move towards thermodynamical equilibrium
but on the contrary: the closer they get to the singularity, the
further away from the said equilibrium the systems depart. By
``system" we mean the apparent cosmic horizon plus the matter and
fields enclosed by it.
\  \\

\noindent At this point it seems suitable to recall that physical
systems tend spontaneously to some equilibrium state compatible
with the constraints imposed on them. This summarizes the
empirical basis of the second law of thermodynamics. Put briefly,
the latter establishes that isolated, macroscopic systems, evolve
to the maximum entropy state consistent with their constraints
\cite{callen}. As a consequence their entropy, $S$, cannot
decrease, i.e., $S' \geq 0$, where the prime means derivative with
respect to the relevant, appropriate variable. Further, $S$ has to
be a convex function of the  said variable, $S'' < 0$, at least at
the last phase of the evolution.
\  \\

\noindent The apparent horizon in FLRW spacetimes is defined as the
marginally trapped surface with vanishing expansion of radius
\cite{bak-rey}
\begin{equation}
\tilde{r}_{A} = 1/\sqrt{H^{2}\, + \, k \, a^{-2}} \, ,
\label{radius}
\end{equation}
\noindent where $a$ and $k$ are the scale factor of the metric and
the spatial curvature index, respectively, and it is widely known
to have an entropy which, leaving aside possible quantum
corrections, is proportional to its area
\begin{equation}
S_{h} \propto {\cal A} = 4 \pi \tilde{r}_{A}^{2} \, ,
 \label{Sh1}
\end{equation}
\noindent which agrees nicely well with the holographic entropy
derived from considerations of the foamy structure  of spacetime
\cite{Ng}. Besides, this horizon appears to be the appropriate
thermodynamic boundary \cite{wang-2006}.
\  \\

\noindent In its turn, the entropy of the fluid enclosed by the
horizon is related to its energy density and pressure by Gibbs'
equation \cite{callen}, namely,
\begin{equation}
T \, dS_{f} = d \left(\rho \, \frac{4 \pi}{3} \tilde{r}_{A}^{3}
\right) \, + \, p \; d \left( \frac{4 \pi}{3}
\tilde{r}_{A}^{3}\right) \, ,
\label{gibbs0}
\end{equation}
\noindent where $T$ stands for the fluid's temperature.
\  \\

\noindent As said above, for the second law to be satisfied the
inequality $dS_{h} \, + \, dS_{f} \geq 0$ must hold at all times,
and $d^{2}(S_{h} \, + \, S_{f}) < 0$ at least at the last stage of
evolution.
\  \\

\noindent Application of this broad idea  to cosmic scenarios
leads to a variety of interesting results. Among others, dark
energy (or some other agent of late acceleration) appears
thermodynamically motivated: both in the case of Einstein gravity
\cite{nd1} and in modified gravity \cite{nd2}. In particular, ever
expanding universes dominated either by radiation or pressureless
matter cannot approach thermodynamic equilibrium at late times.
This is also true for those phantom dominated universes whose
equation of state parameter, $w = p/\rho $, is a constant
\cite{nd1}.
\  \\

\noindent The target of this Letter is to study whether the
universe gets closer and closer to thermodynamic equilibrium as it
approaches a future singularity. We assume Einstein gravity and
provide some general relations in section \ref{general}. Then we
consider three specific FLRW scenarios: the big crunch singularity
(section \ref{bigcrunch}), a sudden singularity (section
\ref{sudden}), and a  ``little rip" singularity (section
\ref{littlerip}). Discussion and final comments are presented in
section \ref{final}.
\section{General relations}
\label{general}
\noindent The field equations for a spatially homogeneous and
isotropic universe are the Friedmann  equation
\begin{equation}
3 \left(H^{2} + \frac{k}{a^{2}}\right) = \frac{3}{\tilde{r}_{A}^{2}} = 8\,\pi\,G\,\rho \, , \qquad \quad
 (k = 0, \pm 1)\, ,
 \label{friedmann}
\end{equation}
and
\begin{equation}
\dot{H}\, = - 4\,\pi\,G\,\left(\rho + p\right)  + \frac{k}{a^{2}}
= - H^{2} - \frac{1}{2 \, \tilde{r}^{2}_{A}}
\left(1+3\frac{p}{\rho}\right) \ .\label{dotH}
\end{equation}
Quite generally we find
\begin{equation}\label{prA}
{\cal A}^{\prime} = 12\pi\frac{\tilde{r}^{2}_{A}}{a}\left(1+\frac{p}{\rho}\right)\ ,
\end{equation}
where the prime denotes the derivative with respect to the scale factor $a$, and
\begin{equation}\label{prprA}
{\cal A}^{\prime\prime} =
36\pi\frac{\tilde{r}^{2}_{A}}{a^{2}}\left(1 +
\frac{p}{\rho}\right) \left[\frac{2}{3}\left(1 +
3\frac{p}{\rho}\right) - \frac{p^{\prime}}{\rho^{\prime}}\right].
\end{equation}
\  \\

\noindent Via the proportionality $S_{h}\propto {\cal A}$
(cf.~(\ref{Sh1})), the relations (\ref{prA}) and (\ref{prprA})
will allow us to obtain information about $S_{h}^{\prime}$ and
$S_{h}^{\prime\prime}$, respectively, for various choices of
$p/\rho$ and $p^{\prime}/\rho^{\prime}$.
\  \\

\noindent To determine the derivatives of $S_{f}$ we must first
discern the temperature evolution of the fluid. From Euler's
relation $n\, T \, s = \rho \, + \, p $, where $n$ and $s$ are the
number density of particles in a comoving volume and the entropy
per particle, respectively, and the conservation equations
$\rho^{\prime} = - 3\left(\rho + p\right)/a$ and $n^{\prime} = -
3n/a$, we find

\begin{equation}\label{prs}
s^{\prime} = p^{\prime} - \left(\rho +p\right)\frac{T^{\prime}}{T}\ .
\end{equation}
Taking into account the perfect fluid condition $s' = 0$, the
temperature behavior is governed by
\begin{equation}
\frac{T'}{T} = \frac{p'}{\rho \, + \, p} \, . \label{Tevol1}
\end{equation}
Straightforwardly one obtains the desired expressions,

\begin{equation}\label{Spr}
S^{\prime}_{f} = 2\pi \frac{\tilde{r}_{A}^{3}}{a}\frac{\rho +p}{T}\left(1 + 3\frac{p}{\rho}\right)
\end{equation}
\noindent and
\begin{equation}\label{prprS}
S^{\prime\prime}_{f} =
2\pi\frac{\tilde{r}_{A}^{3}}{a^{2}}\frac{\rho +
p}{T}\left[12\frac{p}{\rho} \left(1 +
\frac{3}{2}\frac{p}{\rho}\right) -
9\frac{p^{\prime}}{\rho^{\prime}}\left(1 + \frac{p}{\rho}\right) +
\frac{1}{2}\left(1 + 3\frac{p}{\rho}\right)^{2}\right].
\end{equation}
\  \\

\noindent Sections  III to V apply the above set of formulas to
three different cosmological scenarios that harbor a future
singularity.

\section{Big crunch singularity}
\label{bigcrunch}
\noindent Here we focus on the radiation-dominated, spatially
closed (k = +1) FLRW universe and explore the entropy behavior of
the apparent horizon and radiation enclosed by it as the big
crunch draws close. Under such circumstances the temperature gets
so high that matter becomes extremely relativistic and behaves as
radiation, hence $p=\rho/3$ and $\rho a^{4} = $ const.
\  \\

\noindent The scale factor and Hubble function are given by
\begin{equation}
 a(t) = C \, \sqrt{1 \, - \, \left(1 -\frac{t_{s}-t}{C}\right)^{2}}
 \; \quad {\rm and} \; \quad
 H \equiv \frac{\dot{a}}{a} =  -\frac{ \sqrt{C^{2}-a^{2}}}{a^{2}}
 \, , \label{Hubble1}
\end{equation}
where $t_{s}$ is the time at which $a=0$ and $C^{2} \equiv
\textstyle{8 \pi G\over{3}} \, \rho a^{4}$. The horizon area is
${\cal A} = \textstyle{4 \pi \over{C^{2}}}\, a^{4}$. Either by
direct calculation or as special cases from (\ref{prA}) and
(\ref{prprA}) with $p=\rho/3$ it follows that
\begin{equation}
{\cal A}' = \textstyle{16 \pi \over{C^{2}}}\, a^{3} >0 , \; \quad
{\rm and} \; \quad {\cal A}'' = \textstyle{48 \pi\over{C^{2}}}\,
a^{2}
> 0 \, , \label{primeA&2primeA}
\end{equation}
respectively.
\noindent  Thus, the graph of ${\cal A}$  increases and is concave
for all values of the scale factor.
\ \\

\noindent Equation (\ref{Spr}) leads to $ \, S'_{f} \propto a^{2}
> 0 \,$ and equation (\ref{prprS}) implies $ \, S''_{f} \propto a
> 0 \, $. Thus, the total entropy (that of the apparent horizon,
$S_{h}$, plus that of the radiation fluid), fulfills the
generalized second law, $S'_{h} \, + \, S'_{f} > 0$. However,
because $S''_{h} \, + \, S''_{f} > 0$ the system  does not tend to
thermodynamic equilibrium as the big crunch is approached.

\section{Sudden singularity scenario}
\label{sudden}
\noindent In this scenario the singularity occurs at finite time
(say $t_{s}$) and is characterized by the divergence of the
acceleration and pressure while the energy density, scale factor
and Hubble expansion rate remain finite. As the singularity is
approached, the dominant energy condition ($\mid p \mid < \rho$)
is violated  but all other energy conditions are respected.
\  \\

\noindent Let us consider the scale factor of the spatially flat
($k = 0$) FLRW metric as introduced in \cite{jdb1,jdb2}, namely:
\begin{equation}
a(t) = \left( \frac{t}{t_s} \right)^{\alpha} (a_s -1)\, + \, 1 \,
- \, \left(1 - \frac{t}{t_s}\right)^{\beta} \, ,\label{sfactor1}
\end{equation}
where $a_{s} = a(t = t_{s}) > 1$, and $\alpha$ and $\beta$ are
constant parameters lying in the range $(0, 1]$ and $(1, 2)$,
respectively. Obviously, this expression holds for $0 < t <
t_{s}$.
\  \\

\noindent Since
\begin{equation}
\dot{a}(t) = \frac{\alpha}{t_{s}}\, (a_{s} -1) \, \left(
\frac{t}{t_s} \right)^{\alpha -1}\, + \,
\frac{\beta}{t_{s}}\,\left(1 - \frac{t}{t_s}\right)^{\beta -1} > 0
\, , \label{dota1}
\end{equation}
the Hubble function $H = \dot{a}/a$ never becomes negative.
\  \\

\noindent A subsequent derivation yields
\begin{equation}
\ddot{a}(t) = \frac{\alpha}{t_{s}^{2}}\, (\alpha -1 )\, (a_{s} -
1)\, \left( \frac{t}{t_s} \right)^{\alpha -2}\, - \,
\frac{\beta}{t_{s}^{2}}\, (\beta -1) \,\left(1 -
\frac{t}{t_s}\right)^{\beta -2} \, . \label{ddota1}
\end{equation}
\noindent In the limit $t \rightarrow t_{s}$ the first term, both
in $a(t)$ and $\dot{a}(t)$, dominates over the second one while
the latter dominates in the expression for the acceleration which
becomes negative and diverges.  As $\, a \rightarrow a_{s}$, $\, H
\rightarrow H_{s}$ and $\rho \rightarrow \rho_{s} > 0$ where
$a_{s}$, $\, H_{s}$ and $\, \rho_{s}$ are all finite but $p_{s}
\rightarrow \infty$ via the field equation $\, 3 \ddot{a}/a = -4
\pi G (\rho+3p)$.
\  \\

\noindent In view of the above, in the said limit ($t \rightarrow
t_{s}$) we can write
\begin{equation}
\dot{a}(a) = \frac{\alpha}{t_{s}}\, (a_{s} - 1) \, \left(\frac{a
-1}{a_{s}-1}\right)^{(\alpha-1)/\alpha}\, , \qquad H(a) =
\frac{\alpha}{t_{s}} \, (a_{s}-1)^{1/\alpha} \,
\frac{(a-1)^{(\alpha-1)/\alpha}}{a}   \, ,
\label{limitexpressions}
\end{equation}
where we have eliminated the cosmic time in favor of the scale
factor.
\  \\

\noindent The area of the apparent horizon is given by
\begin{equation}
{\cal A} = 4 \pi H^{-2} \simeq 4 \pi \,
\left(\frac{t_{s}}{\alpha}\right)^{2}\, \left(a_{s} - 1
\right)^{-2/\alpha} \frac{a^{2}}{(a-1)^{2(\alpha-1)/\alpha}} \, ,
\label{area1}
\end{equation}
where the second equality holds to leading order only. From (\ref{prA}) one has
\begin{equation}
{\cal A}^{\prime} > 0 \label{primeA}
\end{equation}
always. This corresponds to $H'(a) <0$, which can also be checked
explicitly from the full expression of $H(t)$. Since
$p^{\prime}>0$ (the pressure diverges) and $\rho^{\prime}<0$, both
terms in the square bracket on the right-hand side of equation
(\ref{prprA}) are positive, consequently,
\begin{equation}
{\cal A}^{\prime\prime} > 0 \
\label{2primeA}
\end{equation}
as well.
Both ${\cal A}^{\prime}$ and ${\cal A}^{\prime\prime}$ diverge upon approaching the singularity.
\ \\

 \noindent
Before concluding that in this
sudden singularity scenario the universe departs from
thermodynamic equilibrium we must, as before, examine the
thermodynamic behavior of the gravity source.
\  \\

\noindent From (\ref{Spr}) and (\ref{prprS}) it follows
immediately that with $p>0$, $p^{\prime}>0$  and $\rho^{\prime}<0$
one has both $S^{\prime}_{f}>0$ and $S^{\prime\prime}_{f}>0$. The
inequality $S'_{f}>0$ together with ${\cal A}' > 0$ implies that
the generalized second law, $S'_{h} \, + \, S'_{f} > 0$, is
fulfilled in this scenario. On the other hand, the inequality
$S''_{h} \, + \, S''_{f} > 0$ means that the total entropy is a
concave function of the scale factor and the universe does not
tend to thermodynamic equilibrium.

\section{Little rip scenario}
\label{littlerip}
\noindent The expression ``little rip" was coined as a
contraposition to ``big rip" \cite{caldwell}. In this scenario the
ratio $p/\rho <-1$ but it increases and approaches $\, -1$ as time
goes on. Although sooner or later all bound structures rip apart,
at variance with the usual big rip scenario \cite{caldwell},
neither the energy density nor the scale factor diverge at a
finite time. There is a future singularity but, because the
expansion rate approaches de Sitter, it is pushed to $t
\rightarrow \infty$ \cite{frampton}.
\  \\

\noindent The equation of state of the gravity source is $p =
-\rho - A \rho^{1/2}$, with $A >0$. This alongside the
conservation equation $\rho' = -3(\rho + p)/a$ produces
\begin{equation}
\rho = \rho_{0} \, \left[\frac{3A}{2\sqrt{\rho_{0}}}\, \ln
\left(\frac{a}{a_{0}}\right) \, + \, 1\right]^{2}
\label{endensity}
\end{equation}
\noindent (cf. Eq. (12) in \cite{frampton}). Notice that $\rho$
augments with expansion, a typical feature of phantom  dark
energy, but logarithmically only and diverges just when $a
\rightarrow \infty$. The zero subscript denotes some reference
time, we conveniently take it as the time at which the dark energy
overwhelms all other components (matter, radiation, etc) to the
point that their dynamical influence can be safely ignored.
\  \\

\noindent Because the FLRW metric is spatially flat, we have
${\cal A} = 4 \pi/H^{2} \propto 1/\rho$.
From (\ref{prA}) one verifies ${\cal A}^{\prime} < 0$ and (\ref{prprA})
provides us with ${\cal A}^{\prime\prime} > 0$.
\  \\

\noindent We next study the entropy evolution of the fluid. The
equation of state $p = -\rho - A \rho^{1/2}$ in (\ref{Spr}) yields
$S_{f}^{\prime}> 0$. Since ${\cal A}^{\prime}$ and
$S_{f}^{\prime}$ have opposite signs, a closer look at the
behavior of these quantities as they approach the singularity is
necessary. Explicitly, we have
\begin{equation}\label{prAlr}
{\cal A}^{\prime} = - 12\pi \frac{A}{a H^{2}\rho^{1/2}}
\end{equation}
and
\begin{equation}\label{prSlr}
S_{f}^{\prime} = 2\pi \frac{A\rho^{1/2}}{a H^{3}T}\left(2+3\frac{A}{\rho^{1/2}}\right)\ .
\end{equation}

\noindent To proceed, information about the dependence of the
temperature on $a$ is required. The general law (\ref{Tevol1})
yields
\begin{equation}\label{Tprlr}
\frac{T^{\prime}}{T} = \frac{3}{a}\left(1 + \frac{A}{2\rho^{1/2}}\right)\ .
\end{equation}

\noindent This integrates to
\begin{equation}
\frac{T}{T_{0}} = \left(\frac{a}{a_{0}}\right)^{3} \,
\left[\frac{3A}{2\sqrt{\rho_{0}}}\, \ln
\left(\frac{a}{a_{0}}\right) \, + \, 1 \right] =
\left(\frac{a}{a_{0}}\right)^{3} \,
\left(\frac{\rho}{\rho_{0}}\right)^{1/2}\, . \label{T(a)}
\end{equation}
Consequently,
\begin{equation}\label{prST}
S_{f}^{\prime} = 2\pi A \frac{a_{0}^{3}\rho_{0}^{1/2}}{T_{0}}\,
\frac{2+\frac{3A}{\rho^{1/2}}}{a^{4} H^{3}}\ .
\end{equation}
In approaching the singularity, $\rho$ diverges and the term
$\propto \rho^{-1/2}$ in the second numerator of (\ref{prST}) can
be neglected. Since $H \propto \rho^{1/2}$, one has
$S_{f}^{\prime}\propto 1/(a^{4}\rho^{3/2})$ in the limit of large
values of $a$. In the same limit ${\cal A}^{\prime}$ in
(\ref{prAlr}) behaves as ${\cal A}^{\prime}\propto -
1/(a\rho^{3/2})$. Therefore, as $ a \rightarrow \infty$ the ratio
$\mid {\cal A}'/S'_{f} \mid$ diverges as $a^{3}$ whence $S'_{h} \,
+ \, S'_{f} < 0$ in the long run; i.e., the generalized  second
law is violated as the little rip singularity gets closer and
closer.
\  \\

\noindent Despite we already know that in this scenario no
thermodynamic equilibrium is approached when nearing the
singularity, we shall determine whether the function $S_{h} \, +
\, S_{f}$ is convex or concave in that limit. From (\ref{prprA})
we find that ${\cal A}^{\prime\prime} \propto 1/(\rho^{3/2}a^{2})$
for $a\rightarrow \infty$ while (\ref{prprS}) yields
$S_{f}^{\prime\prime} \propto -1/(\rho^{3/2}a^{5})$ in the
far-future limit. The ratio $\mid {\cal
A}^{\prime\prime}/S^{\prime\prime}_{f} \mid$ diverges with the
same power, $a^{3}$, as the corresponding ratio of the first
derivatives,  i.e., $S''_{h} \, + \, S''_{f}
> 0$ as $a \rightarrow \infty$.

\section{Concluding remarks}
\label{final}
\noindent Macroscopic systems moving by themselves away from
thermodynamic equilibrium is something far removed from daily
experience. This is enshrined in the second law of thermodynamics
that introduces the entropy function and dictates its overall
evolution. In this Letter we have studied the behavior of three
FLRW universes as they draw close to a future singularity. As long
as the entropy concept is related to the apparent cosmic horizon,
none of the three approaches equilibrium.
\  \\

\noindent Against this some comments may be raised: $(i)$ The
proposed universes look too academic; we should not wonder that
unrealistic systems do not comply with thermodynamics. $(ii)$ The
Universe is a very particular and unique system; why should it
obey the thermodynamical laws? $(iii)$ Everyone expects the
breakdown of physical laws at singularities. Then, why should the
second law be an exception?
\  \\

\noindent The first comment seems persuasive; however, take into
account that not so academic models, as is the case of phantom
models with constant $w$, do not approach equilibrium as they near
the singularity \cite{nd1}. Further, it is rather problematic to
draw a dividing line between ``academic" and ``non-academic"
models in cosmology. Regarding the second one, as far as we know,
every macroscopic portion of the Universe fulfills the second law
of thermodynamics. Therefore, there is no compelling argument by
which realistic cosmological models should not fulfill it as well.
As for the third comment, one should not forget that we are
dealing not with the singularities themselves but with the
behavior of the models as they approach their respective
singularities. During the approach physical laws still hold.
\  \\

\noindent From this we may conclude that the three models
considered here are unphysical, at least in the regime nearing the
singularity. The third one, the ``little rip", more particularly
because it violates the generalized second law and it should be
ruled out.
\  \\

\noindent Finally, the fact that these three models do not
approach equilibrium in the last stage of their evolution may be
seen as a feature characterizing future singularities. (This is
shared by the phantom model mentioned above). Nevertheless, since
-for the moment- the existence of counter examples cannot be ruled
out it would be rather premature to assert the general validity of
the said feature in FLRW cosmologies.

\acknowledgments{\noindent DP thanks the ``Departamento de
F\'{\i}sica" of the ``Universidade Federal do Esp\'{\i}rito Santo"
(Vit\'{o}ria), where part of this work was done,  for warm
hospitality and financial support. This research was also
supported by the CNPq and the Ministry of Science and Innovation
under  Grant FIS2009-13370-C02-01, and the ``Direcci\'{o} de
Recerca de la Generalitat" under Grant 2009GR-00164.}


\begin{thebibliography}{99}
\bibitem{stephen-george} S.W. Hawking and G.F.R. Ellis, {\em The Large
Scale Structure of Space-Time} (CUP, Cambridge, 1973).
\bibitem{Bob} R.M. Wald, {\em General Relativity} (The University of Chicago Press, Chicago,
1984).
\bibitem{bouncing} P.J. Steinhardt and N. Turok, Phys. Rev. D
\underline{65}, 126003 (2002); L.R. Abramo, I. Yasuda and P.
Peter, Phys. Rev. D \underline{81}, 023511 (2010).
\bibitem{emergent}
G.F.R. Ellis and R. Maartens, Class. Quantum Grav. \underline{21},
223 (2004); S. del Campo, E. Guendelman, R. Herrera and P.
Labra\~{n}a, JCAP 06(2010)026.
\bibitem{callen} H.B. Callen, {\em Thermodynamics} (J. Wiley, N.Y.,
1960).
\bibitem{bak-rey} D. Bak and S.J. Rey,  Class. Quantum Grav.
\underline{17}, L83 (2000).
\bibitem{Ng} W.A. Christiansen, Y. Jack Ng, and H. van Dam, Phys.
Rev. Lett. \underline{96}, 051301 (2006).
\bibitem{wang-2006} B. Wang, Y. Gong, and E. Abdalla, Phys.
Rev. D. \underline{74}, 083520 (2006).
\bibitem{nd1} Ninfa Radicella and Diego Pav\'{o}n, Gen. Relativ. Grav. (in the press),
arXiv:1012.0474.
\bibitem{nd2} Ninfa Radicella and Diego Pav\'{o}n, Phys. Lett. B
\underline{704}, 260 (2011).
\bibitem{jdb1} J.D. Barrow, Class. Quantum Grav. \underline{21},
L79 (2004).
\bibitem{jdb2} J. D. Barrow, Class. Quantum Grav. \underline{21},
5619 (2004).
\bibitem{caldwell} R. Caldwell {\it et al.}, Phys. Rev. Lett.
\underline{91}, 071301 (2003).
\bibitem{frampton} P.H. Frampton, K.J. Ludwick, and R.J. Scherrer, Phys. Rev. D
\underline{84}, 063003 (2011).
\end{thebibliography}
\end{document}